\documentclass[traditabstract]{aa}

\usepackage{epsfig}

\usepackage{graphicx}
\usepackage[varg]{txfonts}
\usepackage{natbib}

\bibpunct{(}{)}{;}{a}{}{,} % to follow the A&A style

\def\ecs{erg~cm$^{-2}$s$^{-1}$}

\def\bron{4U 1820-30}

\begin{document}

\title{Superexpansion as a possible probe of accretion in \bron}

\titlerunning{Superexpansion as a possible probe of accretion}
\authorrunning{in 't Zand, Homan, Keek \& Palmer}

\author{J. J. M.~in~'t~Zand\inst{1}, J. Homan\inst{2}, L. Keek\inst{3} \& D.M. Palmer\inst{4}}

%\offprints{J.J.M. in 't Zand, email {\tt jeanz@sron.nl}}

\institute{     SRON Netherlands Institute for Space Research, Sorbonnelaan 2,
                3584 CA Utrecht, the Netherlands; {\tt jeanz@sron.nl}
           \and
                Dept. Physics and Kavli Institute for Astrophysics and Space Research,
                Massachusetts Institute of Technology, Cambridge, MA 02139, USA
           \and
                National Superconducting Cyclotron Laboratory, Dept. of Physics \& Astronomy, and Joint Institute for Nuclear Astrophysics, Michigan State University, East Lansing, MI 48824, USA
           \and
                Los Alamos National Laboratory, B244, Los Alamos, NM 87545, USA
          }

\date{\it Accepted on October 2nd, 2012}

\abstract{The ultracompact X-ray binary \bron\ is well known for its
  $\approx$170-d superorbital modulation in X-ray flux and spectrum,
  and the exclusiveness of bursting behavior to the low hard 'island'
  state. In May-June 2009, there was an exceptionally long 51-d low
  state. This state was well covered by X-ray observations and 12
  bursts were detected, 9 with the high-throughput RXTE. We
  investigate the character of these X-ray bursts and find an
  interesting change in their photospheric expansion behavior. At the
  lowest inferred mass accretion rates, this expansion becomes very
  large in 4 bursts and reaches the so-called superexpansion
  regime. We speculate that this is due to the {\em geometry} of
  the inner accretion flow being spherical and a decreasing accretion
  {\em rate}: when the flow geometry nearest to the neutron star is
  spherical and the accretion rate is low, the ram pressure of the
  accretion disk may become too low to counteract that of the
  photospheric expansion. In effect, this may provide a novel means to
  probe the accretion flow. Additionally, we observe a peculiar
  effect: the well-known cessation of X-ray bursts in the high state
  is too quick to be consistent with a transition to stable helium
  burning. We suggest an alternative explanation, that the cessation
  is due to the introduction of a non-nuclear heat source in the
  neutron star ocean.

}

\maketitle

\section{Introduction}
\label{intro}

Low-mass X-ray binaries with neutron stars (NSs) emit X-rays from two
distinct locations: the inner region of the accretion disk, including
perhaps a boundary layer and corona, and the NS. The former is powered
by the liberation of gravitational potential energy and the latter by
thermonuclear burning in the surface layers consisting of accreted
matter. Per nucleon the thermonuclear energy is just up to a few
percent of the gravitational energy and would be hardly
distinguishable if it did not have a different time
scale. Thermonuclear burning is mostly unstable and 'visible' as short
bursts of X-ray radiation while gravitational energy is liberated
throughout the disk on typical time scales of at least a few
weeks. X-ray bursts are characterized by few-second rises and decays
at least ten times as slow, and a cooling black body spectrum of
temperature k$T<3$~keV. For reviews of X-ray bursts, see \cite{lew93}
and \cite{stroh06}.

Different kinds of thermonuclear X-ray bursts exist. One distinction
that one can make is the presence of photospheric radius expansion
(PRE).  The luminosity produced in some bursts is so powerful that the
Eddington limit is reached and the photosphere is lifted from the NS
surface \citep{gri80,ebi83}. PRE can be recognized by a temporarily
increased emission area of a cooler thermal spectrum and a bolometric
flux that remains approximately constant for the same duration. PRE
happens in about 20\% of the bursts \citep{gal08}.  Usually the
increase in emission area is approximately tenfold. In 1\% of all
cases, however, the increase is much larger: 10$^4$ or more (or,
$>10^2$ in terms of photospheric radius). This is qualified as
'superexpansion' (\citealt{zan10}, see also \citealt{hld+78,jvp90}).

There is an intricate interplay between accretion and burst behavior
\citep[e.g., ][]{fuj81,jvp88,cor03}. Primarily the accretion {\em
  rate} determines the burst recurrence time: the faster the accretion
and ergo fuel supply, the faster the bursts recur. At the highest
accretion rates, thermonuclear burning becomes stable and bursts
disappear.  A second order effect involves the hydrogen abundance: a
high abundance may incur a burning regime where stable hydrogen
burning co-exists with unstable helium burning.

The aforementioned effects are observationally obvious.  It is to be
expected, but harder to prove observationally, that there is also a
dependence on accretion {\em geometry}. Non-isotropic accretion on the
NS surface will change the local (or specific) accretion rate and will
influence the ignition conditions \citep[e.g.,][]{bil98} because those
may really depend on the specific rather than global rate. A very nice
illustration of this is the recently discovered transient IGR
J17480--2446 \citep[e.g.,][]{cav11,lin12}.

\begin{figure}
\vspace{-2cm}
\centerline{\includegraphics[width=\columnwidth,angle=0]{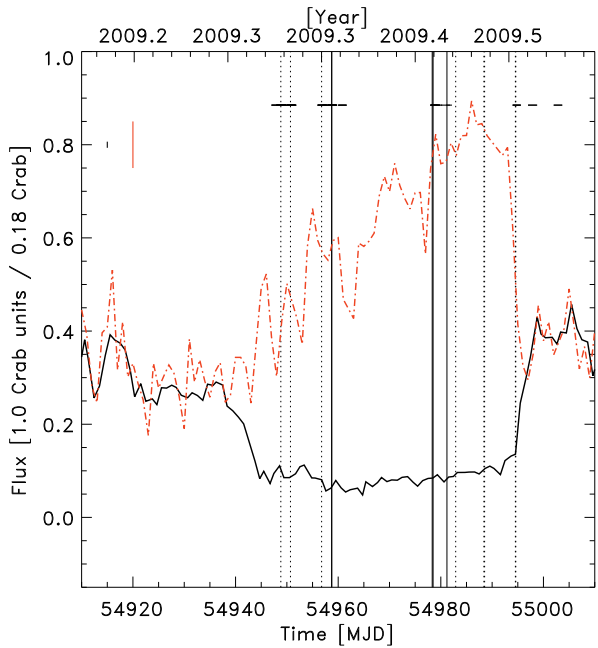}}
\vspace{-3cm}
\caption{The evolution of the 3-12 keV photon flux relative to that of
  the Crab source (black curve; RXTE-ASM data) and the 15-50 keV
  photon flux with respect to 0.18 times that of the Crab source (red
  dash-dotted curve; Swift-BAT data). Vertical lines indicate the
  bursts detected, solid for those with superexpansion and dashed for
  those without. Three bursts are indistinguishable because they have
  too short recurrence times to be visible on this scale (see
  Table~\ref{tabbursts}). The solid horizontal lines indicate the days
  that the PCA observed \bron\ under ObsID 94090. The solid vertical
  lines in the top left corner indicate the 2$\sigma$ uncertainties.
\label{figoutburst}}
\end{figure}

In conclusion, accretion parameters (rate, geometry, composition)
determine burst parameters such as the recurrence time
\citep[e.g.,][]{cor03}, peak luminosity and duration \citep[and, ergo,
  fluence;][]{gal08}, PRE and the amplitude of burst oscillations
\citep{muno04,wat12}.  Conversely, thermonuclear bursts may influence
the accretion.  The luminosity of bursts is sometimes so high that it
may affect the accretion flow for a brief period. The influence may be
dramatic.  \cite{kuu09} discuss an X-ray burst from Cen X-4 which
appears to have turned the accretion disk into a hot state, although
it appears at odds with accretion disk theory. Other examples of
influence are bursts that cool the accretion disk corona
\citep{chen11} and bursts with superexpansion, such as the superburst
from \bron\ \citep{stro02,bal04,kee12} and a number of bursts from
(candidate) ultracompact X-ray binaries \citep{zan10}. During these
bursts, episodes exist when the flux is lower than the pre-burst
accretion flux. This implies that accretion was stopped or covered by
the optically thick shell. Better yet, strong factor-of-2 fluctuations
were recently observed in the decay of several burst light curves that
are very probably due to structural disturbances of the accretion disk
surface \citep{zan11,pal11,deg12}.

Because of the influence of superexpansion on the accretion flow, it
makes sense to study superexpansion as a function of accretion.
Unfortunately, superexpansion is usually a rare phenomenon which makes
such a study difficult. However, in 2009 a series of observations was
done on \bron\ when it was gradually changing its accretion rate and
presumably geometry. Simultaneously, a series of bursts was observed
with different levels of photospheric expansion. We here report our
findings of these observations.

\begin{table*}
\begin{center}
\caption[]{Specifications of all detected bursts detected during the
  long low state (\# 1-12) and 2 bursts that have the smallest hard
  color value in the RXTE archive (\# A-B; see
  Fig.~\ref{figcc}). Numbers between parentheses refer to the
  uncertainty in the last digits.\label{tabbursts}}
\begin{tabular}{rllllllccccccc}
\hline\hline \\
Id. & Instr.  & Date   & Time     & MJD         & RXTE OBSID     & PCUs & SC & HC & \multicolumn{2}{c}{Fluence} & Decay & Pers.flux$^4$ & EED$^5$ \\
    &         &        &          &             &                &      &    &    & \multicolumn{2}{c}{(10$^{-7}$~erg~cm$^{-2}$)} & time$^3$ & (10$^{-9}$~erg& (s)\\
    &         &        & (UT)     &             &                &      &    &    & meas.$^1$ & extrap.$^2$ & (s) & cm$^{-2}$s$^{-1}$) &\\
\hline
 1 & PCA      & Apr 27, 2009 & 19:42:36 & 54948.82125 & 94090-01-01-02 & 2      & 1.92 & 1.08 & 3.2(2) & 4.1 & 3.3(2) & 4.01(3) & 0.70 \\ %1
 2 & PCA      & Apr 29, 2009 & 16:52:04 & 54950.70283 & 94090-01-01-05 & 1,2    & 1.96 & 1.14 & 3.3(2) & 4.3 & 3.4(1) & 3.93(3) & 0.90 \\ %2
 3 & PCA      & May 05, 2009 & 18:35:37 & 54956.77474 & 94090-01-02-03 & 2,4    & 2.05 & 1.19 & 3.4(2) & 4.5 & 3.2(1) & 4.11(3) & 0.90 \\ %5
 4 & PCA\&BAT & May 07, 2009 & 17:45:34 & 54958.73999 & 94090-01-02-02 & 2      & 2.04 & 1.20 & 3.4(2) & 4.3 & 3.4(2) & 4.04(3) & 1.30 \\ %6
 5 & PCA      & May 27, 2009 & 07:42:57 & 54978.32149 & 94090-01-04-00 & 2      & 2.08 & 1.24 & 3.5(2) & 5.0 & 3.7(2) & 4.75(4) & 1.68 \\ %7
 6 & PCA      & May 27, 2009 & 11:52:39 & 54978.49490 & 94090-01-04-01 & 2      & 2.03 & 1.25 & 3.6(2) & 5.2 & 3.6(2) & 4.81(4) & 1.55 \\ %8
 7 & PCA      & May 30, 2009 & 04:29:42 & 54981.18730 & 94090-01-05-00 & 0,2,4  & 2.12 & 1.24 & 3.6(1) & 4.6 & 3.5(1) & 5.00(2) & 1.40 \\ %10
 8 & BAT      & May 31, 2009 & 21:38:16 & 54982.90157 & 		   &   & & &&&&                                         & \\
 9 & BAT      & Jun 06, 2009 & 09:28:19 & 54988.39465 & 		   &   & & &&&&                                         & \\
10 & XRT\&BAT & Jun 06, 2009 & 12:43:37 & 54988.53029 & 		   &   & & &&&&                                         & \\
11 & PCA      & Jun 12, 2009 & 12:49:14 & 54994.53420 & 94090-02-01-00 & 2      & 1.92 & 1.12 & 3.2(2) & 4.2 & 3.1(2) & 6.73(4) & 0.59 \\ %11
12 & PCA      & Jun 12, 2009 & 14:42:45 & 54994.61303 & 94090-02-01-00 & 1,2    & 1.95 & 1.09 & 3.2(1) & 3.6 & 3.5(1) & 6.73(4) & 0.41 \\ %12 
\hline                                                                                                                          
A  & PCA      & Sep 29, 2004 & 10:29:03 & 53277.43685 & 90027-01-03-05 & 0,2,3,4& 1.74 & 0.97 & 3.5(2) & 4.6 & 3.2(1) & 4.28(2) & 0.61 \\
B  & PCA      & Mar 06, 2011 & 18:32:11 & 55626.77236 & 96090-01-01-020& 1,2    & 1.74 & 0.92 & 2.4(2) & 2.9 & 2.7(2) & 4.90(2) & 0.10 \\
\hline\hline
\end{tabular}
\end{center}
$^1$ This is the fluence as measured under the exponential decay
fitted to the time profile of the bolometric flux after the peak, plus
the peak flux times the time interval between the peak flux and the
start of the precursor; $^2$ This is the fluence as measured under the
exponential decay of the bolometric flux, backwards extrapolated to
the start of the precursor; $^3$ The e-folding decay time as
determined from the bolometric flux time history from the peak value
on; $^4$ 2-60 keV flux from fit with comptonized model to all data
from the relevant observation.  $^5$ Superexpansion equivalent
duration (see text)
\end{table*}

\bron\ is the source in which the X-ray burst phenomenon was
discovered by \cite{gri76}.  It is a persistently bright low-mass
X-ray binary (LMXB), viewed at an inclination angle below roughly
50$^{\rm o}$ \citep{and97, bal04}. It is located in the globular
cluster NGC~6624 at a distance of $8.4\pm0.6$~kpc \citep{val04} and is
extraordinary since it has the shortest orbital period known for any
LMXB: 11.4 min \citep{stella87}. The orbital period puts it in the
ultracompact X-ray binary regime \citep{nrj86}. The donor stars in
this group distinguish themselves as being hydrogen-deficient dwarf
stars. The donor in \bron\ is probably a helium white dwarf
\citep{rap87}. From an analysis of burst energetics, \cite{cum03}
  conclude that the hydrogen abundance $X<0.1$. \bron\ shows a
superorbital modulation in the X-ray flux with a period of
$\approx$170~d \citep{pried84,chou01,far09}. The flux resides mostly
in a high state, except for few-day periods every $\approx170$ d when
the (bolometric) flux decreases by a factor of approximately
2. \cite{zdz07b} studied the {\em orbital} modulation as a function of
{\em superorbital} phase and found that the orbital amplitude is up to
a factor of 2 lower during the superorbital low state (or island
state), and that the orbital peak phase shifts. This suggests a
different geometry of X-ray emitting and scattering structures on the
accretion disk between both states. The nature of the $\approx170$~d
periodicity is speculated to be the result of the influence of a third
massive body, see \cite{chou01} and \cite{zdz07a}, which modulates the
orbital eccentricity of the ultracompact system and, thus, the mass
accretion rate.

\begin{figure}
\centerline{\includegraphics[width=\columnwidth,angle=0]{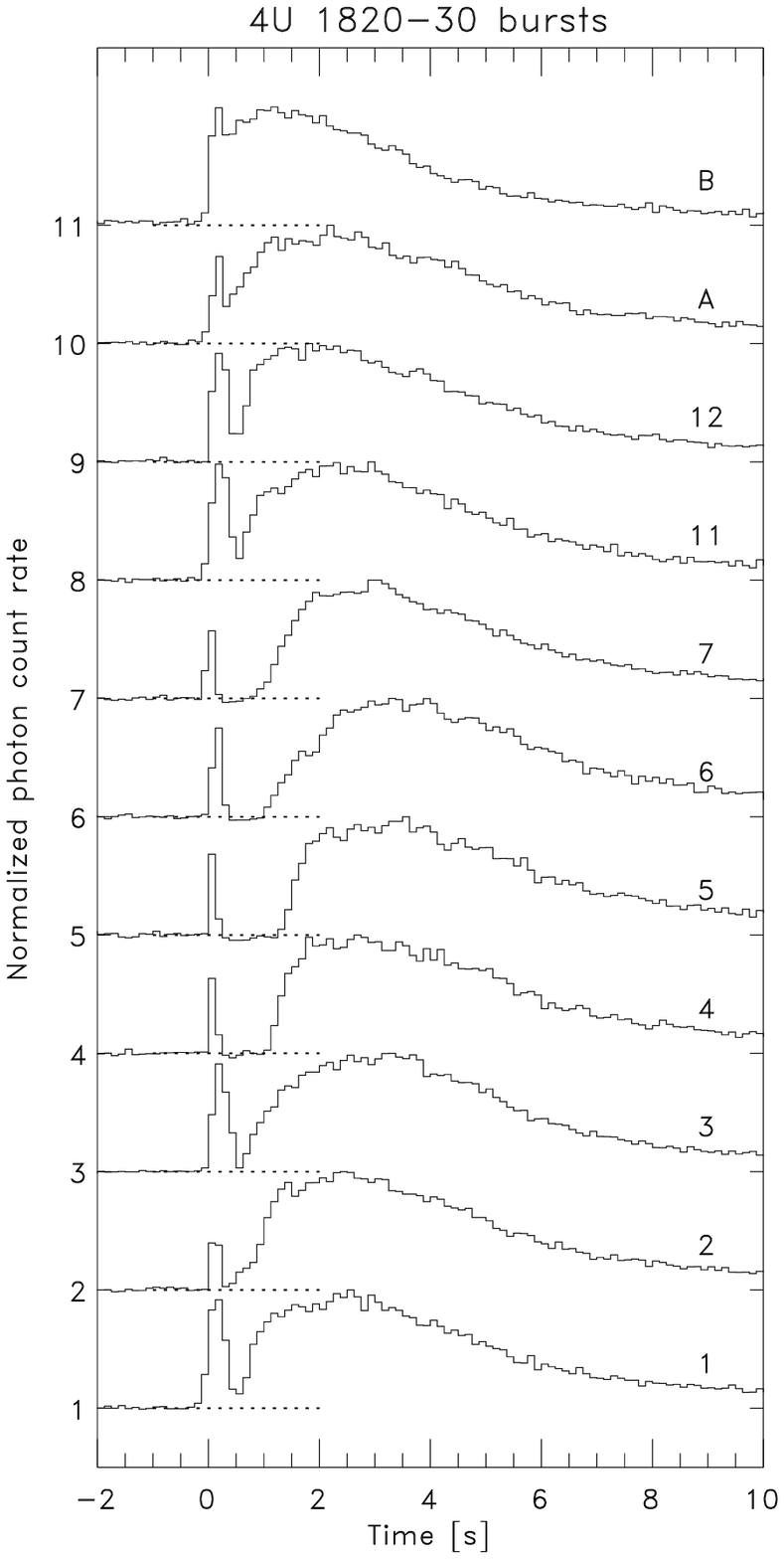}}
\caption{Measured photon rate time histories for the 9 PCA bursts,
  from bottom to top chronologically, supplemented with the 2 PCA
  bursts that are lowest in the color-color diagram
  (Fig.~\ref{figcc}). The subtracted background level is measured from
  a 150 s time interval prior to the burst and the curves have been
  normalized to the peak rate. For guidance, the zero level has been
  indicated by a dotted line.
\label{figlc}}
\end{figure}

It is only during the low states that the source shows X-ray
bursts. This is attributed to the difference in accretion rate: in the
high state thermonuclear burning is stable and not distinguishable
from the accretion radiation while in the low state it is unstable and
visible as X-ray bursts \citep[e.g.,][]{chou01}.  \bron\ is the only
confirmed ultracompact hydrogen-deficient X-ray binary to have
exhibited carbon-fueled superbursts \citep[][see also
  \citealt{bal04}]{stro02,zan11b}.

The subject of our study is the bursting behavior during the longest
low state ever observed in \bron, lasting 51 days in May-June 2009.

\section{Observations} 
\label{obs}

Figure~\ref{figoutburst} shows two light curves of the long low
state. The 2-12 keV RXTE-ASM data \citep[black; for a description of
  the ASM, see][]{lev96} show the typical low/high state behavior. At
MJD 54938 the flux sharply decreases from 0.30 Crab units to 0.10 Crab
five days later. This endures until MJD 54994 (51 days later) when
there is a sharp increase lasting 4 days and reaching 0.40
Crab. During the low state, the flux shows a slow bowl-shaped change
with a minimum halfway through of 0.06 Crab. The bowl is a little
skewed, as though there is a slow increase over a time scale longer
than the low state.

The 15-50 keV BAT data \citep[red curve, as obtained from the
  Swift/BAT Hard X-ray Transient Monitor web site\footnote{{\tt\scriptsize URL
      swift.gsfc.nasa.gov/docs/swift/results/transients/}}; for an
  instrument description, see][]{bar05} show more or less opposite
behavior, with slowly monotonically increasing fluxes during the ASM
low state, from 0.05 Crab to 0.16 Crab at the end of the ASM low
state.

Twelve bursts were detected during the long low state with three
instruments. They are listed in Table~\ref{tabbursts}. Three were
detected with BAT, of which one triggered a slew that resulted in a
detection of the next burst with XRT and BAT. Nine bursts were
detected with the Proportional Counter Array (PCA). Three pairs of
bursts had short wait times of 4.2 hr (bursts 5 and 6), 3.3 hr (bursts
9 and 10) and 1.9 hr (bursts 11 and 12). Interestingly, the last burst
was detected on the same day as the low state ended with the sharp
increase in ASM flux (Fig.~\ref{figoutburst}). The moment of change
was not covered by PCA observations.

The nine bursts detected with the high-throughput PCA on RXTE provide
much better statistical quality than the others. Therefore, we focus
on these. During the low state, \bron\ was observed 27 times for a
total of 134 ks with the RXTE PCA (ObsID 94090).
Figure~\ref{figoutburst} shows the coverage.

The PCA \citep{jah06} consists of 5 proportional counter units (PCUs)
that are sensitive between 2 and 60 keV and combine to give an
effective area of 6000~cm$^2$ at 6 keV. The spectral resolution is
about 18\% full width at half maximum at 6 keV and the time resolution
of the data products for \bron\ is typically 125 $\mu$s. It is rare
that all PCUs are active at the same time, particularly at late times
in the mission, but the central PCU number 2 (counting from 0) is
almost always on. Table~\ref{tabbursts} indicates which PCUs were on
per burst. A wide variety of data products can be requested from the
PCA. For our analyses, we used standard-1 (0.125~s resolution and 1
channel for 2-60 keV photons) and standard-2 data (16~s resolution and
128 channels) and event mode data {\tt E\_125us\_64M\_0\_1s} (125
$\mu$s resolution and 64 energy channels).

Fig.~\ref{figlc} shows the light curves of all PCA bursts as drawn
from the standard-1 data of all operational PCUs.  It is immediately
obvious that for bursts 4, 5, 6 and 7 the photon flux drops below the
pre-burst level, resulting in the appearance of a precursor implying
superexpansion. The photosphere expands while emitting at the
Eddington limit. The increase in emission area is compensated by such
a strong decrease in radiation temperature that the thermal spectrum
shifts out of the PCA bandpass. Additionally, the accretion radiation
is cut short by the expanding shell, either sweeping the inner
accretion disk or covering the X-ray emitting part of the disk. We
note that the XRT burst (number 10) is the first after burst 7, after
six days, with sufficient statistical quality to see superexpansion if
it were present, but none was detected. The 4 superexpansion bursts
are annotated by solid vertical lines in Fig.~\ref{figoutburst} (three
pairs of bursts are indistinguishable in the plot because they occur
within a few hrs).

We include in our analysis 2 other bursts from the total of 17
detected by the PCA in its 16 year mission \citep[for a discussion of
  14 of these, see][]{kee12}, because they are the bursts with the
least amount of expansion. The top two light curves in Fig.~\ref{figlc}
are of these 2 bursts.

\section{Analysis}
\label{ana}

\subsection{Time-resolved burst spectral analysis}

\begin{figure}[t]
\centerline{\includegraphics[width=\columnwidth,angle=0]{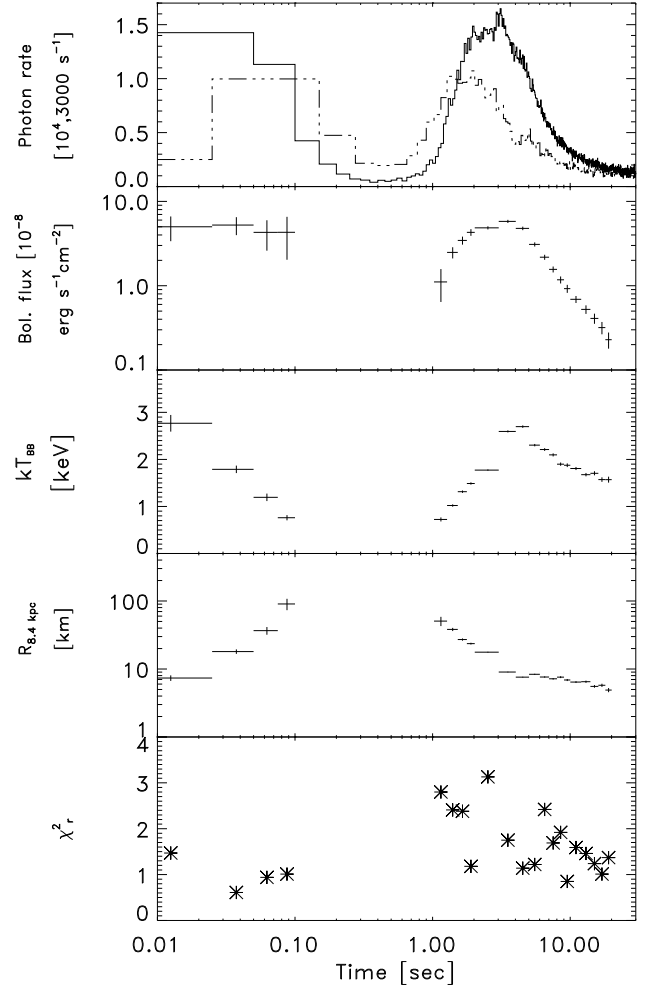}}
\caption{Time-resolved spectroscopy of burst number 7. The top panel
  shows the measured full-bandpass photon count rates for the PCA
  Xenon layers (solid line; units of 10$^4$ s$^{-1}$) and propane
  layers (dashed line; units of 3000 s$^{-1}$). The propane layer has
  a somewhat softer response and therefore better tracks the lower
  energy photons \citep[e.g.,][]{kee12}. 
\label{figspburst7}}
\end{figure}

\begin{figure}[t]
\centerline{\includegraphics[width=0.8\columnwidth,angle=270]{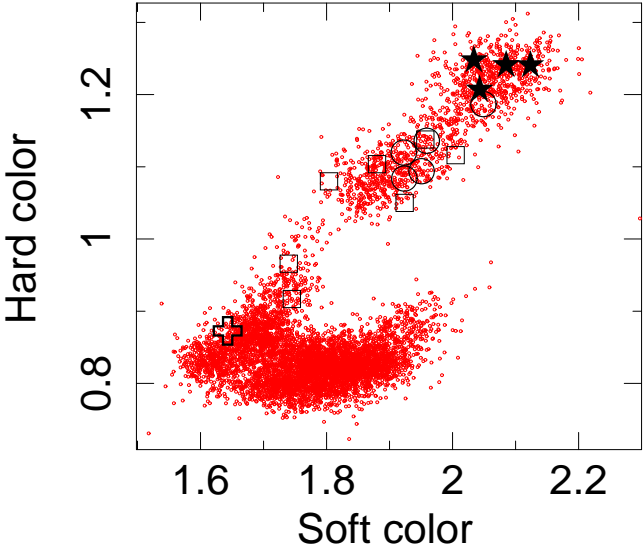}}
\caption{Color-color diagram of all PCU2 measurements of
  \bron\ between March 31, 1999, and March 31, 2011. Colors are
  defined as in \cite{hom10}: soft color is the count rate in PCU2
  channels 10-17 to 6-9 (corresponding to ranges 4.1-7.4 and 2.5-4.1
  keV) and hard color is channels 24-40 to 18-23 (17.1-9.9 and 7.4-9.9
  keV).  Each point is for 128 s of data.  Burst data have been
  excluded.  The stars refer to the colors of the 128-s data stretches
  immediately before the 4 superexpansion bursts, the open circles to
  those of the other 5 bursts in the low state, squares to all other
  ordinary bursts and the large cross to the superburst. One ordinary
  burst is not included here. The burst recurrence time is 2.5 times
  larger at the group of circles with coordinates (1.95,1.09) than at
  the group of stars with coordinates (2.05,1.25).
\label{figcc}}
\end{figure}

We divided each of the 11 PCA bursts in a number of time bins for
which we extracted spectra. The time bins were chosen such as to be
able to follow spectral changes promptly. A spectrum of the non-burst
emission was extracted from standard-2 data in a time interval between
160 and 16 s prior to each burst. The non-source background (cosmic
diffuse emission or particle-induced) was calculated for each
extracted spectrum through {\tt pcabackest} version 3.6 and taken into
account in subsequent analyses. The non-burst source emission was
modeled by a simple spectral shape (thermal bremsstrahlung) that
adhered to the data. Response matrices were calculated for each
observation using {\tt pcarsp} version 11.7.1. Allowance was made for
a systematic error of 1\% per channel.

The burst spectra were modeled by a Planck function for black body
radiation leaving free the temperature and emission area. Allowance
was made for the non-burst emission to vary by including the spectral
model as obtained for the pre-burst interval, with all parameters
fixed, and leaving free its normalization. From the fitted Planck
function, bolometric fluxes were calculated. For PRE phases, it is to
be expected that the shape of the non-burst emission will change. We
do not model this and keep it fixed. Therefore, the black body
parameters are expected to be biased for these phases and we do not
draw conclusions from them. We did not include low-energy absorption
in our models, because the column density ($N_{\rm H}\approx1\times
10^{21}$~cm$^{-2}$; \citealt{kuu03,cos12}) is too low to be noticeable
with the PCA.  Fig.~\ref{figspburst7} shows an example of the fit
results, for a burst that was measured with 3 PCUs (number 7, see
Table~\ref{tabbursts}).

The e-folding decay time $\tau$ was measured by fitting to the time
history of the bolometric flux, beyond the peak, an exponential decay
function. The 'precursor time' was determined as the time between the
peak of the flux and the start of the precursor. The 'gap time' was
estimated by investigating the light curve and estimating the time
between the end of the precursor decay and the rise of the main burst
phase. The burst fluence was determined by combining the integral of
the fitted exponential decay function with the assumed fluence before
the peak flux which is simply determined by the peak flux times the
precursor time. It should be noted that this measures only the nuclear
energy that is transformed into radiation. The nuclear energy may also
transform into kinetic and gravitational energy of expelled
material. A rough estimate of the energy lost to expelled material may
be obtained by back-extrapolating the exponential decay to the time of
the start of the precursor. We find that this amounts to between 10
and 40\% of the (presumed) total nuclear energy. The values are
largest for the four superexpansion bursts, mostly because the
precursor times are longest there.  The results of the measurements
are in Table~\ref{tabbursts} and Fig.~\ref{figburpar}.

There is little variation in burst fluence and duration. The values
vary between 3.2 and 3.6$\times 10^{-7}$~erg~cm$^{-2}$, a 12\% range
with a relative $1\sigma$ uncertainty of about 5\%, and is consistent
with being constant ($\chi^2=11.4$ with 8 d.o.f. for a fit with a
constant). The weighted average is $(3.39\pm0.05)\times
10^{-7}$~erg~cm$^{-2}$. For a $8.4\pm0.6$~kpc distance, the fluence
translates to a total energy of $E_{\rm burst}=(2.9\pm0.4)\times
10^{39}$~erg. If this is due to a pure layer of helium, with a
$3\alpha$ energy production of $E_{\rm nuc}=5.84\times
10^{17}$~erg~g$^{-1}$, the ignition column depth is $y_{\rm
  ign}\approx E_{\rm burst}/4\pi R^2 E_{\rm nuc}=(4.0\pm0.7)\times
10^8$~g~cm$^{-2}$ for a canonical NS radius of 10 km. The upper limit
on the variation in fluence (10\%) is, unfortunately, not accurate
enough to constrain the hydrogen abundance in a manner suggested by
\cite{cum03}.

\subsection{Spectral modeling of accretion-powered radiation}
\label{sec32}

To put the data in perspective, we show in Figure~\ref{figcc} the
commonly used \citep[e.g.,][]{hom10} color-color diagram of all PCA
data of \bron\ obtained after March 22, 1999. We exclude earlier data
(including one burst) because early gain changes of the detectors
preclude straightforward inference of colors. The data show two
branches, commonly referred to as the banana state (lower branch) and
island state (upper branch) \citep{has89}. All low states are
equivalent to the island state. Also shown in Fig.~\ref{figcc} are the
data points for the 17 PCA bursts detected in this period. These are
taken from the 128-s intervals immediately prior to the
bursts. Obviously, all bursts occur in the island state. Most notably,
the superexpansion bursts occur in the extreme upper part of the
island state.

It is difficult to find an unambiguous physical model for the spectra
of the accretion radiation, or 'persistent spectrum' as we shall call
it. Low-resolution LMXB 2-30 keV spectra are most often modeled with a
combination of (disk) black body and comptonization components,
representative of the thermal radiation from the accretion disk, NS
and/or boundary layer, and the comptonization in a corona in the
neighborhood of that disk. The geometry of the corona is uncertain. We
tested this model against the spectra of the complete observations for
each burst, employing PCA standard product spectra. In particular, we
employed a simple black body component and a Comptonized component
according to the prescription by \cite{tit94}. In XSPEC they are
identified by model names {\tt 'bbody'} and {\tt 'comptt'}. We find
that the evidence for the black body component is very uncertain in
these data and that including it in the model degrades the accuracy
with which comptonization parameters can be measured, due to strong
coupling between the two components (both are very broad continua
without unambiguous distinctive significantly detected spectral
features). Therefore, we choose to ignore the black body
component. This implies somewhat worse fits, but the $\chi^2_\nu$
values remain fairly close to acceptable - between 1 and 3.9. The
spectra are well represented by the comptonization component within a
level of 3\% per channel. We here present the fitted parameters, with
errors that have been multiplied by a factor of $\sqrt{\chi^2_\nu}$
which would be equivalent with a fit with $\chi^2_\nu=1$. The
resulting 2-60 keV fluxes are provided in Table~\ref{tabbursts} and
Fig.~\ref{figburpar}, plasma temperatures and optical depths (for a
disk geometry) are provided in Fig.~\ref{figburpar}. We note that the
energy of the seed photons is always far below the lower bandpass
threshold and, thus, not constrained by the data. The spectral results
show that the plasma temperature and optical depth varies
significantly by a factor of 3 for 2-60 keV fluxes that are
comparable.

For a 8.4 kpc distance, the 2-60 keV fluxes translate to a 2-60 keV
luminosity range of $3.3$ to $5.7\times10^{37}$~\ecs. For a canonical
10~km/1.4~M$_\odot$ NS this is equivalent to a mass accretion rate of
2.8 to 4.8 $\times 10^{-9}$~M$_\odot$~yr$^{-1}$ or a specific mass
accretion rate of 1.4 to 2.4$\times 10^4$ g~s$^{-1}$cm$^{-2}$.

\subsection{Correlation study}

We introduce a new empirical parameter to measure the amount and
duration of expansion: the equivalent expansion duration (EED). The
basis for this is the background-subtracted normalized light curve as
shown for all bursts in Fig.~\ref{figlc}. The EED is similarly defined
as the equivalent width of absorption lines in spectra: it is the loss
of photons due to the expansion, expressed in seconds that it would
take to emit those photons at the level of the peak flux. The EED is a
function of the bandpass and the effective area function, but this is
of no consequence here because we use the same instrument in its full
bandpass throughout. When superexpansion occurs, the light curve drops
to negative values. This will push the EED to even higher values.

\begin{figure}
\centerline{\includegraphics[width=\columnwidth,angle=0]{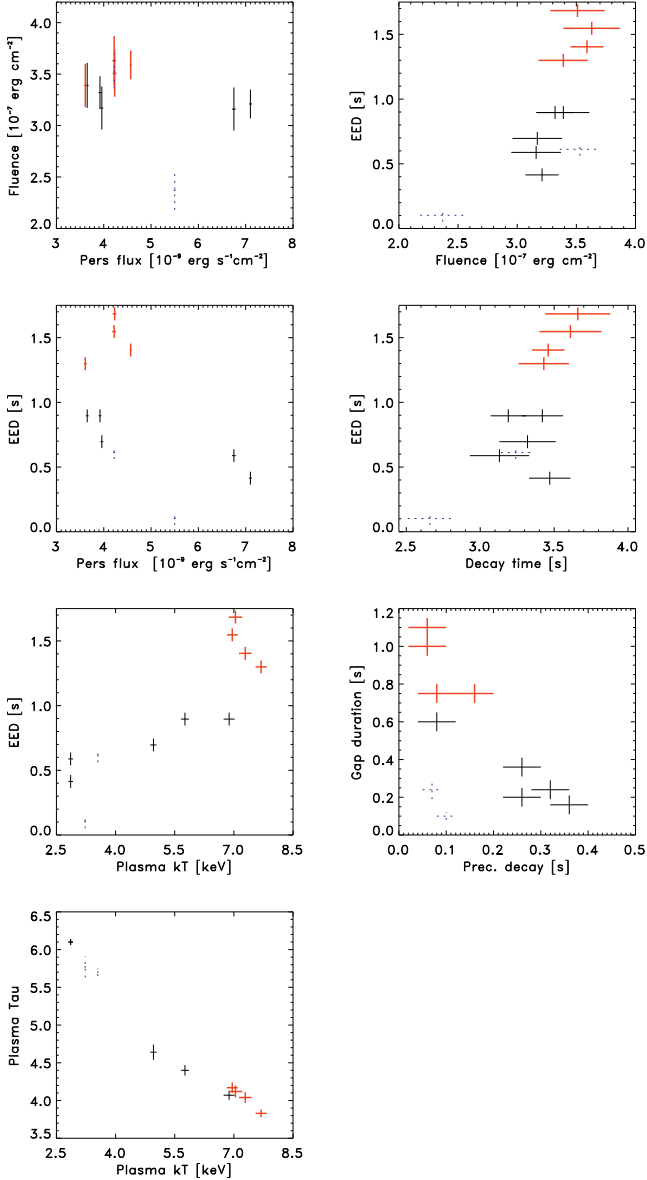}}
\vspace{0cm}
\caption{Diagrams of burst parameters versus persistent emission
  parameters (left column, except bottom plot) and burst parameters
  versus burst parameters (right column). The plotted fluence is the
  measured fluence (see Table~\ref{tabbursts}). The dashed blue data
  points refer to bursts A and B (top 2 light curves in
  Fig.~\ref{figlc}). The 4 superexpansion bursts are plotted with
  thick red lines. For additional information a diagram is shown
  between plasma temperature and optical depth (bottom left plot).
\label{figburpar}}
\end{figure}

We plot various burst parameters against other burst parameters as
well as accretion parameters in Fig.~\ref{figburpar}.  The only strong
one-to-one correlation exist between EED and the shape of the
pre-burst spectrum as measured through the comptonizing plasma
parameters.

\section{Discussion}
\label{secdis}

The data can be summarized as follows:
\begin{list}{\leftmargin=0.4cm \itemsep=0cm \parsep=0cm \topsep=0cm}
\item
\item[$\bullet$] the fluence and decay rate of all bursts are equal
  within 6\%, except for burst B.  The helium content is presumably
  close to 100\%, see below. This implies that the difference in
  photospheric expansion is unlikely to be the result of a difference
  in nuclear energy generated or ignition depth;
\item[$\bullet$] the amount of expansion is strongly correlated with
  the position of \bron\ on the color-color diagram in the sense that
  the 4 bursts with the strongest expansion (superexpansion) are at
  one extreme end of the island state and those with the smallest
  expansion at the opposite end. Equivalently, the EED is strongly
  correlated with the comptonization parameters;
\end{list}

\noindent
This strongly suggests that the accretion environment determines the
amount of expansion exhibited by thermonuclear X-ray bursts. The
question is: in what way? Is it related to the {\em rate} of
accretion, the {\em geometry} of the accretion, or both?

All changes in the persistent radiation are thought to be driven by
changes in accretion rate. By how much does the accretion rate change?
It is well known that fluxes do not directly translate to accretion
rates.  It appears to be true only on order-of-magnitude scales. One
can think of various reasons why there is no 1-to-1 correspondence
between flux and accretion rate: the measured persistent flux is not
bolometric, the flux may have changing anisotropy factors and the
radiation efficiency of transforming gravitational to radiation energy
may change.  Fortunately, our data provide an independent means of
measuring $\dot{M}$ changes. The three recurrence times measured for
X-ray bursts from \bron\ (see Sect.~\ref{obs}), in combination with
the burst fluences, indicate that the accretion rate at burst 12
(without superexpansion) is 2.5$\pm0.2$ times larger than at burst 6
\citep[with super expansion; provided that all nuclear burning is
  unstable and that the nuclear and gravitational energy liberated per
  accreted nucleon is constant; i.e., the burst alpha parameter,
  see][is constant]{lew93}. For comparison, the 2-60 keV persistent
flux ratio is only 1.4$\pm$0.1 over these bursts. Note that this ratio
difference supports the pure helium accretion hypothesis: the pure
helium model in \cite{cum03} predicts that $\dot{M}$ should be 1.5 to
2 times higher than the value derived from the X-ray flux to have
agreement with burst recurrence times and energies in \bron.

Changes in spectral shape (as measured via for instance color in
Fig.~\ref{figcc} or plasma temperature in Fig.~\ref{figburpar}) are
generally attributed to changes in the geometry of the inner accretion
flow, probably as a result of changes in accretion rate. The low/hard
and high/soft states are very common in variable LMXBs, disregarding
the nature of the accretor \citep[NS or black hole; see reviews by
][respectively]{mvk06,rem06}. These distinct states are usually
accompanied by distinct Fourier spectra.

PCA timing and spectral studies of \bron\ have been the subject of
previous papers.  \cite{zha98} and \cite{blo00} report on twin kHz
quasi-periodic oscillations (QPOs) and find that they, from the lower
left banana state to the island state, gradually move in tandem to
frequencies twice as small \citep[see also][]{alt05}. In the banana
state, the frequencies saturate \citep[but see][]{men02}. \cite{blo00}
modeled the PCA spectra by a combination of a black body and
comptonization component and find that in the island state the plasma
temperature increases and the optical depth decreases. We observe the
same trend further into the island state (Fig.~\ref{figburpar}). This
behavior is consistent with an interpretation that the QPOs originate
in the inner accretion disk \citep[e.g.,][]{mvk06} and that the inner
edge of the accretion disk moves outward (recedes) from the lower left
banana to the upper island state and that the cleared region is filled
by a non-thermal more spherical advection-dominated (i.e., with less
radiative efficiency) accretion flow. This is supported by the
findings of \cite{zdz07b} mentioned in Sect.~\ref{intro}. 

We note that we have searched for QPOs in the new data, but were able
to find only a single, broad QPO around 540 Hz in the combined power
spectra of the three observations with a soft color lower than 2.0
(i.e. roughly middle of the island state). A comparison with the power
spectra of 4U 1820-33 analyzed in \cite{alt05} suggests that this QPO
is the upper kHz QPO. We note that our data are statistically less
constraining than those of Altamirano et al. (2005) due to a smaller
number of active PCUs. The other five observations, with soft colors
higher than 2.0, extend to more extreme parts of the island state, in
which the inner accretion disk edge possibly moves out even
further. No high frequency QPO is found in these observations,
although indications for a broad (Q~0.6) feature around 280 Hz are
seen.

Expansion into the circumstellar medium will become easier when the
ram pressure of the accretion flow becomes smaller. This may happen if
the accretion rate decreases or if the geometry changes. The latter
may occur, for example, if the accretion is channeled through a disk
instead of spherically. A large part of the solid angle as seen from
the NS will then open up and allow free expansion. However, all bursts
that we observe occur during the island state, when the inner flow is
thought to be spherical all the time and the inner edge of the
accretion disk is thought to be at some distance from the NS. If the
frequency is Keplerian in nature, a factor of 2 decrease in frequency
(see above), implies a factor of $2^{2/3}=1.6$ increase in radius from
banana to island state.

The change in expansion factor is, therefore, most likely to be
directly related to the accretion rate. In the following we attempt to
find quantitative support for this. The ram pressure of free falling
spherical accretion can be expressed as \citep[][page 135]{fra92}
\begin{eqnarray}
P_{\rm ram, acc} & = & \rho v_{\rm ff}^2 
\label{eqn1}
\end{eqnarray}
with $\rho$ the density of the infalling matter and $v_{\rm ff}$ the free-fall
velocity 
\begin{eqnarray}
v_{\rm ff} & = & \left( 2GM/R \right)^{1/2}
\end{eqnarray}
The density can be derived from the continuity equation
\begin{eqnarray}
4 \pi R^2 \rho v & = & \dot{M}
\end{eqnarray}
so that Eqn. \ref{eqn1} becomes
\begin{eqnarray}
P_{\rm ram, acc} & = & \frac{(2GM)^{1/2}}{4\pi} \; \dot{M} \; R^{-5/2} \nonumber \\
               & = & 1.5\times 10^{12} \; \dot{M} \; R^{-5/2} \; {\rm dyne~cm^{-2}}
\end{eqnarray}
for a 1.4~$M_\odot$ NS, with $\dot{M}$ in g~s$^{-1}$ and $R$ in cm.

The ram pressure of a shell leaving the NS is
\begin{eqnarray}
P_{\rm ram, shell} & = & \rho \; v_{\rm expand}^2 \nonumber \\
                 & = & \frac{y_0}{l} \; \left(\frac{R}{R_{\rm NS}}\right)^{-2}  \; v_{\rm expand}^2
\end{eqnarray}
where $y_0$ is the initial column thickness of the shell, $l$ the
geometrical thickness and $v_{\rm expand}$ the expansion velocity of
the shell. 

The shell will be able to leave the NS surface ($R=R_{\rm NS}$) if
$P_{\rm ram, shell}>P_{\rm ram, acc}$ or if
\begin{eqnarray}
\dot{M} < 6.7\times 10^{-13} \; \frac{y_0}{l} \; v_{\rm expand}^2 \; R_{\rm NS}^{5/2}
\label{eqn2}
\end{eqnarray}
Once the shell has left the NS, it depends on how the shell thickness
evolves whether its ram pressure will remain above that of the
accretion flow.  For a constant shell thickness it does, but for a
shell thickness that grows stronger than $R^{0.5}$ it does not.

What values can be applied for the parameters in Eq.~\ref{eqn2}?  We
use the initial rate of increase in the black body radius (see
Fig.~\ref{figspburst7}) as a rough estimate of the expansion velocity:
$v_{\rm expand}\sim 10^3$~km~s$^{-1}$.  The shell column thickness
$y_0$ may be derived from the superexpansion duration
\citep[see][]{zan10}. It takes about 1 s for the shell to become
transparent (at a column thickness of 1~g~cm$^{-2}$) due to dilution
that goes as $R^{-2}$.  During that time the shell has expanded to a
radius of $v_{\rm expand}\times t\sim10^3$~km. Back extrapolating the
dilution implies that $y_0\sim 10^4$~g~cm$^{-2}$. This is $10^{-4}$
times the ignition of $y_{\rm ign}\approx 4\times
10^8$~g~cm$^{-2}$. For the geometrical layer thickness one may use the
pressure scale height of a helium atmosphere which is $\sim33$ cm
\citep[e.g.,][]{pir07}. For $R_{\rm NS}=10^6$~cm, Eq.~\ref{eqn2}
predicts that the shell will be able to lift off if $\dot{M} \la 3
\times 10^{-5}$ $M_\odot$~yr$^{-1}$. This is a very loose constraint
which is fullfilled under any practical circumstance, 10$^4$ times
larger than measured for \bron.  This may be due to the relative
arbitrariness of the chosen parameter values. For instance, $l$ may
increase substantially during the expansion and $v_{\rm expand}$ may
evolve.  If $l$ is $\sim$100 m instead of 33 cm and $v_{\rm expand}$ 3
times smaller, the constraint in Eq.~\ref{eqn2} would be similar to
the observed $\dot{M}$. Furthermore, we have used a very simple
approach (isotropic spherical free-fall accretion, isotropic shell
expansion, constant shell velocity and thickness, ignoring radiation
pressure, and ignoring GR corrections). The conclusion is that a
quantitative verification of our proposed scenario of the
superexpansion behavior in \bron\ needs detailed modeling which is
outside the scope of the present paper.

The constant ignition depth would imply that the expelled shell
thickness is similar in all bursts, but that the amount of distance
traveled by the shell differs. Usually, it does not travel far,
perhaps less than 100 km, before falling back. Only for the
superexpansion bursts does it travel far enough to move out of the
X-ray band. This is supported by the fact that there is always
evidence for a shell, even if there is no superexpansion: all bursts
from \bron\ show a drop in flux within 1 s from the burst onset.  This
idea is supported by the slower decay of the initial spike for bursts
without superexpansion. The shell is moving slower and does not reach
high altitudes (i.e., in Fig.~\ref{figlc} the spikes drop the fastest
in the superexpansion case).

We note that the superburst in \bron\ also has superexpansion while it
is on the opposite side of the island state in the color-color diagram
(Fig.~\ref{figcc}). A new analysis of this event \citep{kee12} shows
that the energy contained in the precursor is larger than those in
ordinary bursts from \bron, which may be explained by a shock induced
by the carbon detonation and reaching the photosphere. Therefore, the
driving of the expansion occurs differently from that in ordinary
bursts from \bron\ and the shell ram pressure is likely to be
different.

\section{A note on the lack of X-ray bursts in the banana state}

As an addendum, we address a peculiar phenomenon in our measurements:
X-ray bursting activity seems to react more rapidly to changes in the
accretion-induced radiation than expected from crustal heating.  The
first burst after the flux arrives in the low state is detected within
5 days. The last burst before the persistent flux returns back to the
high state is detected only two days before. No bursts were detected
in PCA measurements within the next 8 days with a total exposure of
10.4 ks while the recurrence time between the two last bursts was 6.8
ks s, nor with Swift observations. We made a similar but more
significant measurement in 2011, when we followed up a PCA burst
detection with a 1.2~d long continuous Chandra observation starting
1.3 d after the burst. The jump to the high state occurred in the 1.3
d time frame between both observations.  No burst was detected in RXTE
data and Chandra data within 2.5~d after the burst with a total
exposure time of 1.4~d.  In the same time frame there were 0.3~d of
exposure time with Swift-BAT, without any X-ray burst detection. We
note that a similarly sudden stop of bursting activity was observed in
IGR J17473-2721 \citep{che11}.

The data suggest that bursting activity stops as suddenly as the low
state does. This is peculiar. In the common picture of \bron\ if
$X<0.03$ \citep{cum03}, the bursting activity is driven by the
temperature of the crust. The crust is heated through pycnonuclear
reactions and electron-capture reactions \citep[e.g.,][]{gup07,hae08}
whose rates are a direct function of the accretion rate.  However, the
thermal time scale of the crust is of order months
\citep[e.g.,][]{bro09}, so there should be a delay between a change in
the accretion rate and one in the bursting activity. Therefore, the
prompt lack of X-ray bursts in the high state of \bron, and vice
versa, is not explained by stabilization of helium burning by crustal
heating.

An alternative explanation might be that in the banana state the inner
edge of the accretion disk is probably closer to the NS, see
Sect.~\ref{secdis}. Depending on the inclination angle and thickness
of the inner disk and/or boundary layer, this is perhaps sufficient to
block our view of the NS and make X-ray bursts invisible. A test of
this hypothesis would be the detection of partly obscured X-ray bursts
with a few-hours recurrence times. This has never been observed in 4U
1820-30, despite abundant coverage throughout the history of X-ray
astronomy. Furthermore, X-ray bursts have been observed in banana
states of other sources \citep[e.g.,][]{klis90,mun00}. Therefore, this
explanation is unlikely.

\cite{che11} proposed another explanation for the sudden stop of
bursting activity in IGR J17473-2721: the occurrence of a superburst
during a data gap. However, given the long recurrence time of
superbursts (of the order of 1 yr for persistent sources and longer
for transients such as IGR J17473-2721), this appears to us an
unlikely scenario.

A more promising explanation seems that during the high state the
primary heating of the flash layers does not occur by the crust, but
by a heat source at a shallower depth. A thermal time scale of $\la
1$~d applies to column depths of $\la 10^{12}$~g~cm$^{-2}$
\citep{bro09} which is in the NS ocean and outside the crust. The heat
source is most likely not of nuclear origin. For example, it cannot be
CNO burning, because there is no significant abundance of hydrogen in
the accreted material. \cite{ino10} suggest that for accretion rates
in excess of a few percent of the Eddington limit, rotational heating
may stabilize helium burning. In this model, the braking of a
hypersonic azimuthal flow in the accretion disk by the NS atmosphere
results in an additional heat source. Perhaps the accretion flow
geometry in the banana state of this source allows for more efficient
transfer of angular momentum from the disk to the NS envelope,
boosting this heat source and quenching the bursts. Quantitative
assessment of this hypothesis requires 2-dimensional multi-zone
simulations of nuclear burning, primarily as a function of NS spin,
accretion composition and flow geometry. Unfortunately, the NS spin
has not been measured yet for 4U 1820-30.

\acknowledgements

We thank Nevin Weinberg (MIT), Ed Brown (MSU) and Tullio Bagnoli
(SRON) for useful discussions. JZ and LK are members of an
International Team in Space Science on type-I X-ray bursts sponsored
by the International Space Science Institute (ISSI) in Bern and thank
ISSI for the hospitality during part of this work. We are grateful to
the anonymous referee for constructive comments that helped improve
the paper. LK is supported by the Joint Institute for Nuclear
Astrophysics (JINA; grant PHY08-22648), a National Science Foundation
Physics Frontier Center. This paper utilizes data from the
Multi-INstrument Burst ARchive (MINBAR). RXTE/ASM and Swift/BAT
results were provided by the RXTE/ASM and Swift/BAT team,
respectively.

\bibliographystyle{aa} \bibliography{references}

\begin{thebibliography}{60}
\expandafter\ifx\csname natexlab\endcsname\relax\def\natexlab#1{#1}\fi

\bibitem[{{Altamirano} {et~al.}(2005){Altamirano}, {van der Klis},
  {M{\'e}ndez}, {Migliari}, {Jonker}, {Tiengo}, \& {Zhang}}]{alt05}
{Altamirano}, D., {van der Klis}, M., {M{\'e}ndez}, M., {et~al.} 2005, \apj,
  633, 358

\bibitem[{{Anderson} {et~al.}(1997){Anderson}, {Margon}, {Deutsch}, {Downes},
  \& {Allen}}]{and97}
{Anderson}, S.~F., {Margon}, B., {Deutsch}, E.~W., {Downes}, R.~A., \& {Allen},
  R.~G. 1997, \apjl, 482, L69+

\bibitem[{{Ballantyne} \& {Strohmayer}(2004)}]{bal04}
{Ballantyne}, D.~R. \& {Strohmayer}, T.~E. 2004, \apjl, 602, L105

\bibitem[{{Barthelmy} {et~al.}(2005){Barthelmy}, {Barbier}, {Cummings},
  {Fenimore}, {Gehrels}, {Hullinger}, {Krimm}, {Markwardt}, {Palmer},
  {Parsons}, {Sato}, {Suzuki}, {Takahashi}, {Tashiro}, \& {Tueller}}]{bar05}
{Barthelmy}, S.~D., {Barbier}, L.~M., {Cummings}, J.~R., {et~al.} 2005, Space
  Science Reviews, 120, 143

\bibitem[{Bildsten(1998)}]{bil98}
Bildsten, L. 1998, in The many faces of neutron stars, ed. A.~Alpar,
  L.~Buccheri, \& J.~van Paradijs, NATO ASI (Kluwer, Dordrecht), 419

\bibitem[{{Bloser} {et~al.}(2000){Bloser}, {Grindlay}, {Kaaret}, {Zhang},
  {Smale}, \& {Barret}}]{blo00}
{Bloser}, P.~F., {Grindlay}, J.~E., {Kaaret}, P., {et~al.} 2000, \apj, 542,
  1000

\bibitem[{{Brown} \& {Cumming}(2009)}]{bro09}
{Brown}, E.~F. \& {Cumming}, A. 2009, \apj, 698, 1020

\bibitem[{{Cavecchi} {et~al.}(2011){Cavecchi}, {Patruno}, {Haskell}, {Watts},
  {Levin}, {Linares}, {Altamirano}, {Wijnands}, \& {van der Klis}}]{cav11}
{Cavecchi}, Y., {Patruno}, A., {Haskell}, B., {et~al.} 2011, \apjl, 740, L8

\bibitem[{{Chen} {et~al.}(2011){Chen}, {Zhang}, {Torres}, {Zhang}, {Li},
  {Kretschmar}, \& {Wang}}]{chen11}
{Chen}, Y.-P., {Zhang}, S., {Torres}, D.~F., {et~al.} 2011, \aap, 534, A101

\bibitem[{{Chenevez} {et~al.}(2011){Chenevez}, {Altamirano}, {Galloway}, {in't
  Zand}, {Kuulkers}, {Degenaar}, {Falanga}, {Del Monte}, {Evangelista},
  {Feroci}, \& {Costa}}]{che11}
{Chenevez}, J., {Altamirano}, D., {Galloway}, D.~K., {et~al.} 2011, \mnras,
  410, 179

\bibitem[{{Chou} \& {Grindlay}(2001)}]{chou01}
{Chou}, Y. \& {Grindlay}, J.~E. 2001, \apj, 563, 934

\bibitem[{{Cornelisse} {et~al.}(2003){Cornelisse}, {in 't Zand}, {Verbunt},
  {Kuulkers}, {Heise}, {den Hartog}, {Cocchi}, {Natalucci}, {Bazzano}, \&
  {Ubertini}}]{cor03}
{Cornelisse}, R., {in 't Zand}, J.~J.~M., {Verbunt}, F., {et~al.} 2003, \aap,
  405, 1033

\bibitem[{{Costantini} {et~al.}(2012){Costantini}, {Pinto}, {Kaastra}, {in't
  Zand}, {Freyberg}, {Kuiper}, {M{\'e}ndez}, {de Vries}, \& {Waters}}]{cos12}
{Costantini}, E., {Pinto}, C., {Kaastra}, J.~S., {et~al.} 2012, \aap, 539, A32

\bibitem[{{Cumming}(2003)}]{cum03}
{Cumming}, A. 2003, \apj, 595, 1077

\bibitem[{{Degenaar} {et~al.}(2012){Degenaar}, {Altamirano}, \&
  {Wijnands}}]{deg12}
{Degenaar}, N., {Altamirano}, D., \& {Wijnands}, R. 2012, The Astronomer's
  Telegram, 4219, 1

\bibitem[{{Ebisuzaki} {et~al.}(1983){Ebisuzaki}, {Hanawa}, \&
  {Sugimoto}}]{ebi83}
{Ebisuzaki}, T., {Hanawa}, T., \& {Sugimoto}, D. 1983, \pasj, 35, 17

\bibitem[{{Farrell} {et~al.}(2009){Farrell}, {Barret}, \& {Skinner}}]{far09}
{Farrell}, S.~A., {Barret}, D., \& {Skinner}, G.~K. 2009, \mnras, 393, 139

\bibitem[{{Frank} {et~al.}(1992){Frank}, {King}, \& {Raine}}]{fra92}
{Frank}, J., {King}, A., \& {Raine}, D. 1992, {Accretion power in astrophysics,
  2nd ed.} (Cambridge University Press)

\bibitem[{{Fujimoto} {et~al.}(1981){Fujimoto}, {Hanawa}, \& {Miyaji}}]{fuj81}
{Fujimoto}, M.~Y., {Hanawa}, T., \& {Miyaji}, S. 1981, \apj, 247, 267

\bibitem[{{Galloway} {et~al.}(2008){Galloway}, {Muno}, {Hartman}, {Psaltis}, \&
  {Chakrabarty}}]{gal08}
{Galloway}, D.~K., {Muno}, M.~P., {Hartman}, J.~M., {Psaltis}, D., \&
  {Chakrabarty}, D. 2008, \apjs, 179, 360

\bibitem[{Grindlay {et~al.}(1976)Grindlay, Gursky, Schnopper, Parsignault,
  Heise, Brinkman, \& Schrijver}]{gri76}
Grindlay, J., Gursky, H., Schnopper, H., {et~al.} 1976, ApJ, 205, L127

\bibitem[{{Grindlay} {et~al.}(1980){Grindlay}, {Marshall}, {Hertz},
  {Weisskopf}, {Elsner}, {Ghosh}, {Darbro}, {Sutherland}, \& {Soltan}}]{gri80}
{Grindlay}, J.~E., {Marshall}, H.~L., {Hertz}, P., {et~al.} 1980, \apjl, 240,
  L121

\bibitem[{{Gupta} {et~al.}(2007){Gupta}, {Brown}, {Schatz}, {M{\"o}ller}, \&
  {Kratz}}]{gup07}
{Gupta}, S., {Brown}, E.~F., {Schatz}, H., {M{\"o}ller}, P., \& {Kratz}, K.-L.
  2007, \apj, 662, 1188

\bibitem[{{Haensel} \& {Zdunik}(2008)}]{hae08}
{Haensel}, P. \& {Zdunik}, J.~L. 2008, \aap, 480, 459

\bibitem[{{Hasinger} \& {van der Klis}(1989)}]{has89}
{Hasinger}, G. \& {van der Klis}, M. 1989, \aap, 225, 79

\bibitem[{Hoffman {et~al.}(1978)Hoffman, Lewin, Doty, \& et~al.}]{hld+78}
Hoffman, J., Lewin, W., Doty, J., \& et~al. 1978, ApJ (Letters), 221, L57

\bibitem[{{Homan} {et~al.}(2010){Homan}, {van der Klis}, {Fridriksson},
  {Remillard}, {Wijnands}, {M{\'e}ndez}, {Lin}, {Altamirano}, {Casella},
  {Belloni}, \& {Lewin}}]{hom10}
{Homan}, J., {van der Klis}, M., {Fridriksson}, J.~K., {et~al.} 2010, \apj,
  719, 201

\bibitem[{{in 't Zand} {et~al.}(2011{\natexlab{a}}){in 't Zand}, {Serino},
  {Kawai}, \& {Heinke}}]{zan11b}
{in 't Zand}, J., {Serino}, M., {Kawai}, N., \& {Heinke}, C.
  2011{\natexlab{a}}, The Astronomer's Telegram, 3625, 1

\bibitem[{{in 't Zand} {et~al.}(2011{\natexlab{b}}){in 't Zand}, {Galloway}, \&
  {Ballantyne}}]{zan11}
{in 't Zand}, J.~J.~M., {Galloway}, D.~K., \& {Ballantyne}, D.~R.
  2011{\natexlab{b}}, \aap, 525, A111

\bibitem[{{in 't Zand} \& {Weinberg}(2010)}]{zan10}
{in 't Zand}, J.~J.~M. \& {Weinberg}, N.~N. 2010, \aap, 520, A81

\bibitem[{{Inogamov} \& {Sunyaev}(2010)}]{ino10}
{Inogamov}, N.~A. \& {Sunyaev}, R.~A. 2010, Astronomy Letters, 36, 848

\bibitem[{{Jahoda} {et~al.}(2006){Jahoda}, {Markwardt}, {Radeva}, {Rots},
  {Stark}, {Swank}, {Strohmayer}, \& {Zhang}}]{jah06}
{Jahoda}, K., {Markwardt}, C.~B., {Radeva}, Y., {et~al.} 2006, \apjs, 163, 401

\bibitem[{{Keek}(2012)}]{kee12}
{Keek}, L. 2012, ArXiv e-prints

\bibitem[{{Kuulkers} {et~al.}(2003){Kuulkers}, {den Hartog}, {in 't Zand},
  {Verbunt}, {Harris}, \& {Cocchi}}]{kuu03}
{Kuulkers}, E., {den Hartog}, P.~R., {in 't Zand}, J.~J.~M., {et~al.} 2003,
  \aap, 399, 663

\bibitem[{{Kuulkers} {et~al.}(2009){Kuulkers}, {in 't Zand}, \&
  {Lasota}}]{kuu09}
{Kuulkers}, E., {in 't Zand}, J.~J.~M., \& {Lasota}, J.-P. 2009, \aap, 503, 889

\bibitem[{{Levine} {et~al.}(1996){Levine}, {Bradt}, {Cui}, {Jernigan},
  {Morgan}, {Remillard}, {Shirey}, \& {Smith}}]{lev96}
{Levine}, A.~M., {Bradt}, H., {Cui}, W., {et~al.} 1996, \apjl, 469, L33+

\bibitem[{{Lewin} {et~al.}(1993){Lewin}, {van Paradijs}, \& {Taam}}]{lew93}
{Lewin}, W.~H.~G., {van Paradijs}, J., \& {Taam}, R.~E. 1993, Space Science
  Reviews, 62, 223

\bibitem[{{Linares} {et~al.}(2011){Linares}, {Altamirano}, {Chakrabarty},
  {Cumming}, \& {Keek}}]{lin12}
{Linares}, M., {Altamirano}, D., {Chakrabarty}, D., {Cumming}, A., \& {Keek},
  L. 2011, ArXiv e-prints

\bibitem[{{Mendez}(2002)}]{men02}
{Mendez}, M. 2002, ArXiv Astrophysics e-prints

\bibitem[{{Muno} {et~al.}(2000){Muno}, {Fox}, {Morgan}, \& {Bildsten}}]{mun00}
{Muno}, M.~P., {Fox}, D.~W., {Morgan}, E.~H., \& {Bildsten}, L. 2000, \apj,
  542, 1016

\bibitem[{{Muno} {et~al.}(2004){Muno}, {Galloway}, \& {Chakrabarty}}]{muno04}
{Muno}, M.~P., {Galloway}, D.~K., \& {Chakrabarty}, D. 2004, \apj, 608, 930

\bibitem[{{Nelson} {et~al.}(1986){Nelson}, {Rappaport}, \& {Joss}}]{nrj86}
{Nelson}, L.~A., {Rappaport}, S.~A., \& {Joss}, P.~C. 1986, \apj, 304, 231

\bibitem[{{Palmer}(2011)}]{pal11}
{Palmer}, D. 2011, The Astronomer's Telegram, 3663, 1

\bibitem[{{Piro} \& {Bildsten}(2007)}]{pir07}
{Piro}, A.~L. \& {Bildsten}, L. 2007, \apj, 663, 1252

\bibitem[{{Priedhorsky} \& {Terrell}(1984)}]{pried84}
{Priedhorsky}, W. \& {Terrell}, J. 1984, \apjl, 284, L17

\bibitem[{{Rappaport} {et~al.}(1987){Rappaport}, {Ma}, {Joss}, \&
  {Nelson}}]{rap87}
{Rappaport}, S., {Ma}, C.~P., {Joss}, P.~C., \& {Nelson}, L.~A. 1987, \apj,
  322, 842

\bibitem[{{Remillard} \& {McClintock}(2006)}]{rem06}
{Remillard}, R.~A. \& {McClintock}, J.~E. 2006, \araa, 44, 49

\bibitem[{{Stella} {et~al.}(1987){Stella}, {Priedhorsky}, \&
  {White}}]{stella87}
{Stella}, L., {Priedhorsky}, W., \& {White}, N.~E. 1987, \apjl, 312, L17

\bibitem[{{Strohmayer} \& {Bildsten}(2006)}]{stroh06}
{Strohmayer}, T. \& {Bildsten}, L. 2006, {New views of thermonuclear bursts}
  (Compact stellar X-ray sources), 113--156

\bibitem[{{Strohmayer} \& {Brown}(2002)}]{stro02}
{Strohmayer}, T.~E. \& {Brown}, E.~F. 2002, \apj, 566, 1045

\bibitem[{{Titarchuk}(1994)}]{tit94}
{Titarchuk}, L. 1994, \apj, 434, 570

\bibitem[{{Valenti} {et~al.}(2004){Valenti}, {Ferraro}, \& {Origlia}}]{val04}
{Valenti}, E., {Ferraro}, F.~R., \& {Origlia}, L. 2004, \mnras, 351, 1204

\bibitem[{{van der Klis}(2006)}]{mvk06}
{van der Klis}, M. 2006, {Rapid X-ray Variability}, ed. {Lewin, W.~H.~G.~\& van
  der Klis, M.} (Cambridge University Press), 39--112

\bibitem[{{van der Klis} {et~al.}(1990){van der Klis}, {Hasinger}, {Damen},
  {Penninx}, {van Paradijs}, \& {Lewin}}]{klis90}
{van der Klis}, M., {Hasinger}, G., {Damen}, E., {et~al.} 1990, \apjl, 360, L19

\bibitem[{{van Paradijs} {et~al.}(1990){van Paradijs}, {Dotani}, {Tanaka}, \&
  {Tsuru}}]{jvp90}
{van Paradijs}, J., {Dotani}, T., {Tanaka}, Y., \& {Tsuru}, T. 1990, \pasj, 42,
  633

\bibitem[{{van Paradijs} {et~al.}(1988){van Paradijs}, {Penninx}, \&
  {Lewin}}]{jvp88}
{van Paradijs}, J., {Penninx}, W., \& {Lewin}, W.~H.~G. 1988, \mnras, 233, 437

\bibitem[{{Watts}(2012)}]{wat12}
{Watts}, A.~L. 2012, ArXiv e-prints

\bibitem[{{Zdziarski} {et~al.}(2007{\natexlab{a}}){Zdziarski},
  {Gierli{\'n}ski}, {Wen}, \& {Kostrzewa}}]{zdz07b}
{Zdziarski}, A.~A., {Gierli{\'n}ski}, M., {Wen}, L., \& {Kostrzewa}, Z.
  2007{\natexlab{a}}, \mnras, 377, 1017

\bibitem[{{Zdziarski} {et~al.}(2007{\natexlab{b}}){Zdziarski}, {Wen}, \&
  {Gierli{\'n}ski}}]{zdz07a}
{Zdziarski}, A.~A., {Wen}, L., \& {Gierli{\'n}ski}, M. 2007{\natexlab{b}},
  \mnras, 377, 1006

\bibitem[{{Zhang} {et~al.}(1998){Zhang}, {Smale}, {Strohmayer}, \&
  {Swank}}]{zha98}
{Zhang}, W., {Smale}, A.~P., {Strohmayer}, T.~E., \& {Swank}, J.~H. 1998,
  \apjl, 500, L171

\end{thebibliography}

\end{document}